\documentclass[aps,prd,twocolumn,showpacs,groupedaddress,titlepage]{revtex4-1}  

\usepackage{dcolumn}   
\usepackage{bm}        
\usepackage{amssymb}   
\usepackage{mathtools} 
\usepackage{hyperref}
\hypersetup{colorlinks=true,linkcolor=blue,urlcolor=blue,citecolor=blue}
\usepackage{enumerate}
\usepackage{accents}

\newcommand{\rmi}{{\rm i }}
\newcommand{\JMP}{J. Math. Phys. }
\newcommand{\CQG}{Classical Quantum Gravity }
\newcommand{\PRL}{Phys. Rev. Lett. }

\newcommand{\MPLA}{Mod. Phys. Lett. A } 
\newcommand{\GRG}{Gen. Relativ. Gravit. }
\newcommand{\PLB}{Phys. Lett. B }
\newcommand{\PRD}{Phys. Rev. D }

\begin{document}

\title{Gauge connection formulations for general relativity}

\author{Diego Gonz\'alez}
\email[]{dgonzalez@fis.cinvestav.mx}
\affiliation{Departamento de F\'{\i}sica, Cinvestav, Instituto Polit\'ecnico Nacional 2508, San Pedro Zacatenco,
	 07360, Gustavo A. Madero, Ciudad de M\'exico, M\'exico}

\author{Merced Montesinos} 
\email[]{merced@fis.cinvestav.mx}
\affiliation{Departamento de F\'{\i}sica, Cinvestav, Instituto Polit\'ecnico Nacional 2508, San Pedro Zacatenco, 
	07360, Gustavo A. Madero, 
Ciudad de M\'exico, M\'exico}

\date{\today}

\begin{abstract}
We report a new class of $SO(3,\mathbb{C})$ and diffeomorphism invariant formulations for general relativity with either a vanishing or a nonvanishing  cosmological constant, which depends functionally  on a $SO(3,\mathbb{C})$ gauge connection and a complex-valued 4-form via a holomorphic function of the trace of a symmetric $3\times3$ matrix that is constructed from these variables. We present two members of this class, one of which results from the implementation of a method for obtaining action principles belonging to the class. For the case of a nonvanishing cosmological constant, we solve for the complex-valued 4-form and get pure connection action principles.  We perform the canonical analysis of the class. The analysis shows that only the Hamiltonian constraint  is modified with respect to the Ashtekar formulation and that the members of the class have two physical degrees of freedom per space point.
\end{abstract}


\maketitle

\section{Introduction}

The gauge connection formulations for classical general relativity (GR) are relatively new ways of thought that open, in principle, new roads toward the quantization of the gravitational field~\cite{quantum}. In these formulations, the gauge connection is the main structure used to describe the gravitational field, while the metric is a derived object. A remarkable feature that makes them attractive is that the field equations can be substantially simpler than the ones emerging directly from Einstein's original formulation based on the metric field.  Among these approaches we find the one due to Plebanski, who showed that GR can be expressed as a BF theory supplemented with a constraint on the B fields. This formulation involves a $SO(3,\mathbb{C})$ gauge connection, a $SO(3,\mathbb{C})$ 2-form, a $SO(3,\mathbb{C})$ scalar field, and a Lagrange multiplier~\cite{plebanski}. Based on such a work,  Capovilla, Dell, and Jacobson (CDJ) went one step further and presented a pure spin connection formulation for GR without a cosmological constant, where the variables involved are a gauge connection and a scalar density~\cite{CJD1}. The generalization of this formulation to the case including a nonvanishing cosmological constant has shown to be a nontrivial task. Indeed, the same authors  attempted to carry this out~\cite{CJD2}, however, erroneously~\cite{erratum}. A correct action principle involving the cosmological constant was derived later by Capovilla and Jacobson~\cite{CJ3} together with Peldan~\cite{Peldan}, using totally different methods. Unfortunately, this action principle has not been widely applied because of some technical aspects that prevent its handling.  Recently, Krasnov has achieved a different action principle for GR that requires a nonvanishing cosmological constant and that depends functionally on a $SO(3)$ gauge connection only~\cite{Kras1}.  

In this paper, we report a new class of gauge connection formulations for GR, in the Lorentzian signature case, which has as fundamental variables a $SO(3,\mathbb{C})$ gauge connection and a complex-valued 4-form. We explore the members of this class and find a new action principle for GR that works well with or without a cosmological constant. Furthermore, considering some caveats that we will clarify later on, a particular case of this member can be related to the action principle found in Ref.~\cite{erratum}. We also develop a method for constructing the members of the class by integrating certain holomorphic functions. This method is illustrated with a new action principle for GR with a nonvanishing cosmological constant. We also derive pure connection action principles from the class with a nonvanishing cosmological constant, by eliminating the dependence of the action on the auxiliary 4-form. Finally, we develop the canonical analysis of the class and show that the Hamiltonian constraint is modified with respect to the Ashtekar formulation, whereas the Gauss and vector constraints remain unchanged. The class has two complex degrees of freedom per space point.

\section{The class of formulations for general relativity}\label{class}

Let us begin by fixing the notation and convention. The fundamental variables considered in this paper are a $SO(3,\mathbb{C})$ gauge connection $A^i$ with curvature  $F^i=d A^i + \frac{1}{2} {\varepsilon}^i{}_{jk}  A^j  \wedge A^k$ and a nonvanishing complex-valued 4-form $\rho$. The indices $i,j,k=1,2,3$ are raised and lowered with the Kronecker delta $\delta_{ij}$ and ${\varepsilon}_{ijk}$ is the Levi-Civit\`a symbol  (${\varepsilon}_{123}=+1$). The wedge product of forms is denoted by $\wedge$. Now, let us define a complex $3\times3$ matrix $\Psi(A^i,\rho)$, which is a function of $A^i$ and $\rho$ via
\begin{eqnarray}
F^i\wedge F^j +2 \rho X^i{}_k X^{jk}=0, \label{FF-cond}
\end{eqnarray}
where $X\equiv\Psi+(1/3)\Lambda I$. Here,  $\Lambda$ is the usual cosmological constant and $I$ is the identity $3\times3$ matrix.  

From now on we restrict the analysis to configurations such that $X$ is symmetric and nonsingular, as is usual in the pure spin connection formulations for GR~\cite{CJD2,CJ3}. This restriction is needed in order to hold the equivalence between our formulation and Plebanki's equations of motion for GR. Because $X$ is nonsingular, then the symmetric density matrix $\tilde{M}$ of weight 1 and defined by $\tilde{M}^{ij}d^4x=F^i\wedge F^j$ is also nonsingular on account of (\ref{FF-cond}). Since $X$ is symmetric, then (\ref{FF-cond}) becomes  ${\tilde M}=-2 {\tilde \mu} X^2$ where $\rho = {\tilde \mu} d^4x$. The action principles considered in this paper depend on  ${\rm Tr} \Psi \equiv \Psi^{ij} \delta_{ij}$; therefore, we need to compute $X$ from ${\tilde M}=-2 {\tilde \mu} X^2$, next compute $\Psi$, and finally compute ${\rm Tr} \Psi $. The solution for $X$ exists~\cite{book}, but is not unique generically~\cite{book2}. Nevertheless, the action principles introduced below work well for any of these solutions. In the pure spin connection formulations for GR there is a debate about whether an additional criterion must be introduced in order to select just one or whether all the solutions must be allowed to describe GR~\cite{CJ3, Peldan}. 

Having defined $\Psi(A^i,\rho)$, we are ready to give the action principle 
\begin{eqnarray}
S[A^i,\rho]=\int \rho f({\rm Tr}\Psi),\label{Action}
\end{eqnarray}
where $f$ is a given holomorphic function that depends on ${\rm Tr}\Psi$ and that has the same dimensions of the cosmological constant. It is worth pointing out that $\Psi^{ij}$ is a tensor density of weight zero, and hence any function $f$ is also of weight zero and leaves the action (\ref{Action}) correctly  defined. Then it makes sense to consider an arbitrary holomorphic function $f$. However, an arbitrary choice of $f$ may lead to other theories of gravity different from GR. In this paper we emphasize that we only consider functions $f$ such that   (\ref{Action}) is an action for GR. The other possible cases are also interesting and will be treated elsewhere. 

We find that GR emerges from the class of formulations (\ref{Action}) if $f$ is a holomorphic function in a domain $\Omega$ in $\mathbb{C}$ to which zero belongs, and if it satisfies the following properties:
\begin{enumerate}[(i)]
\item The only zero of $f- \frac{1}{2} \left( {\rm Tr}\Psi + \Lambda \right) f'$, as a function of ${\rm Tr}\Psi$, is ${\rm Tr}\Psi=0$;
\item $f'_0\equiv\left. f' \right|_{{\rm Tr}\Psi=0} \neq 0$;
\end{enumerate}
where ``$\prime$" denotes the derivative with respect to ${\rm Tr}\Psi$.  The requirement of conditions (i) and (ii) will be evident below.

Now we prove that given a function $f$ satisfying (i) and (ii), the class of formulations (\ref{Action}) describes GR with and without a cosmological constant. To do this, it is useful to begin by determining the equations of motion of the action (\ref{Action}), in the case in which $f$ is an arbitrary holomorphic function. These equations are given by
\begin{eqnarray}
\delta \rho: && f- \frac{1}{2} \left( {\rm Tr}\Psi + \Lambda \right) f'=0, \label{mot-rho}\\
\delta A^i:&& D\left[-\frac{f'}{2}    (X^{-1})^i{}_j F^j \right] =0,\label{mot-con}
\end{eqnarray}
where $D$ is the covariant derivative with respect to $A^i$. Here we have used our assumption that $X$ is symmetric and nonsingular.  To compute the variations of ${\rm Tr}\Psi$ with respect to $\rho$ and $A^i$, we employ Eq. (\ref{FF-cond}) to get $\rho\delta{\rm Tr}\Psi= -\frac{1}{2} {\rm Tr}X \delta\rho$ and $\rho\delta{\rm Tr}\Psi= -\frac{1}{2} (X^{-1})_{ij} F^i\wedge \delta F^j$, respectively. 

To simplify Eqs. (\ref{mot-rho}) and (\ref{mot-con}), we now consider that $f$ is endowed with the desired properties (i) and (ii). Since $f$ satisfies condition (i), equation of motion (\ref{mot-rho}) implies
\begin{eqnarray}
{\rm Tr}\Psi=0. \label{cond-1}
\end{eqnarray}
Notice that (\ref{cond-1}) is equivalent to the constraint ${\rm Tr}X=\Lambda$, which  also appears in the pure spin connection formulation of Ref.~\cite{CJ3}. We remark that in our approach, as well as in Ref.~\cite{CJ3}, the appropriate value for ${\rm Tr}\Psi$ comes from the equation of motion for the auxiliary field $\rho$, and that a different value of ${\rm Tr}\Psi$ will lead to other theories of gravitation. This is one reason why the field $\rho$ is relevant.

Moreover, note that $X^{ij}$ and $f'$ have weight zero, and hence the term in the argument of the covariant derivative of Eq. (\ref{mot-con})  is actually a well-defined 2-form.  Furthermore, this 2-form must be evaluated at ${\rm Tr}\Psi=0$ because of (\ref{cond-1}), and, at this point, it does not vanish trivially because $f$ satisfies the property (ii). Therefore, the equation of motion (\ref{mot-con}) acquires the form
\begin{eqnarray}
D\Sigma^i=0,   \label{PlebConn}
\end{eqnarray}
in which the 2-form $\Sigma^i$ is given by
\begin{eqnarray}
\Sigma^i=   (X^{-1})^i{}_j F^j. \label{sigma}
\end{eqnarray} 
Note that $f'_0$ does not appear in (\ref{PlebConn}), since it is a nonvanishing dimensionless constant. Now it is clear the role of conditions (i) and (ii).

The remarkable fact is that Eqs. (\ref{cond-1}) and (\ref{PlebConn}) along with the definition (\ref{sigma}) are Einstein's equations for GR. Indeed, combining (\ref{FF-cond}) with (\ref{sigma}), we obtain the Plebanski constraints  
\begin{eqnarray}
\Sigma^i \wedge \Sigma^j + 2 \rho \delta^{ij}=0. \label{PlebSigma}
\end{eqnarray}
This means that the $\Sigma$'s are actually those of the Plebanski formulation for  general relativity. Furthermore, notice that $\Sigma^i \wedge \Sigma_i\neq0$ since $ \rho\neq0$. Then, Eq. (\ref{PlebSigma}) together with the reality conditions  
\begin{eqnarray}
 \Sigma^i \wedge {\overline \Sigma}^j = 0, \hspace{7mm} \Sigma^i \wedge \Sigma_i + {\overline \Sigma}^i \wedge {\overline \Sigma}_i =0,
\end{eqnarray}
imply $\Sigma^i = \rmi \theta^0 \wedge \theta^i - \frac{1}{2} \varepsilon^i{}_{jk} \theta^j \wedge \theta^k$, where $\{ \theta^0 , \theta^i \}$ are four linearly independent real 1-forms and $\rmi$ is the imaginary unit. The overbar  denotes the complex conjugate. We recall that the reality conditions do not come from the formulation (\ref{Action}), but they are introduced by hand in order to relate the complex 2-forms $\Sigma^i$ with real 1-forms~\cite{plebanski,CDJ4}. 

On the other hand, Eq. (\ref{PlebConn}) is a system of 12 linear equations for the 12 unknowns contained in the components of $A^i$. For Plebanski's $\Sigma^i$ this system is nondegenerate and can be solved for $A^i=A^i(\Sigma)$~\cite{GMV}; however, further assumptions are needed to link $A^i$ with a spacetime connection because there are no spacetime geometrical structures involved in the class of gauge connection formulations (\ref{Action}) nor in any other gauge connection formulation~\cite{GMV}. According to Levi-Civit\`a, a spacetime connection is uniquely defined by specifying  its torsion and its action on a spacetime metric. Therefore, the first assumption consists in defining the spacetime metric, which is given by the Urbantke metric in terms of $\Sigma^i$~\cite{urbantke}, namely, $(1/12) \varepsilon_{ijk} \Sigma^i{}_{MI} \Sigma^j{}_{JK} \Sigma^k{}_{LN}  {\tilde \eta}^{IJKL}$, where $I,J,\ldots=0,1,2,3$ and  ${\tilde \eta}^{IJKL}$ is a totally antisymmetric density of weight 1 (${\tilde \eta}^{0123}=1$). A direct calculation using the expression for  $\Sigma^i$  in terms of the real 1-forms shows that this metric turns out to be the Minkowski metric. The second assumption is that the spacetime connection has no torsion. As a result, the connection $A^i$ is the self-dual part of the spin connection (i.e., the Levi-Civit\`a spacetime connection) and Eq. (\ref{PlebConn}) becomes the first Cartan's structure equation with vanishing torsion. This in turn implies that $F^i$ is the self-dual part of the curvature of the spin connection. Next, $F^i$ can be solved from (\ref{sigma}),
\begin{eqnarray}
F^i= \Psi^i{}_j \Sigma^j +\frac{1}{3} \Lambda \Sigma^i , \label{PlebCurv}
\end{eqnarray}
which means that $F^i$ is self-dual as a 2-form. 

Finally, Eqs. (\ref{cond-1}) and (\ref{PlebCurv}) together with the relation between $F^i$ and curvature of the spin connection imply Einstein's equations for GR. Indeed, Eqs. (\ref{cond-1}), (\ref{PlebConn}),  (\ref{PlebSigma}), and (\ref{PlebCurv})  are \textit{exactly} the set of equations of motion of the Plebanski formulation for complex GR with a cosmological constant. More details on the Plebanski formulation can be obtained in Refs.~\cite{GMV,Kras2}. In addition, the trace-free symmetric matrix $\Psi$ becomes the self-dual part of the Weyl tensor.

\section{Example: polynomial functions}

To illustrate how the class of formulations (\ref{Action}) works, we present the simple, but rather significant, case of a quadratic function of the ${\rm Tr}\Psi$. To this end, let us begin by considering the function
\begin{eqnarray}
f =\alpha_1 (2 {\rm Tr}\Psi+\Lambda)+ \alpha_2 ({\rm Tr}\Psi+\Lambda)^2, \label{fquadratic}
\end{eqnarray}
where $\alpha_1$ and $\alpha_2$ are arbitrary constants with appropriate units. Our goal is to demonstrate that, with a specific choice of $\alpha_1$ and $\alpha_2$,  the function (\ref{fquadratic})  is holomorphic on a certain region $\Omega$ containing zero and satisfying conditions (i) and (ii). The first requirement is satisfied because (\ref{fquadratic}) is a polynomial function, and hence holomorphic in the whole complex plane. Conditions (i) and (ii) are satisfied if  $\alpha_1\neq0$ and $\alpha_1+\alpha_2\Lambda\neq0$, respectively. Note that these conditions are fulfilled for $\Lambda=0$ as well as for $\Lambda\neq0$. Indeed if, for instance, $\Lambda=0$, then both conditions reduce to $\alpha_1\neq0$. In view of this, we can conclude that the new action principle (\ref{Action}) with $f$ given by (\ref{fquadratic}), where $\alpha_1\neq0$ and $\alpha_1+\alpha_2\Lambda\neq0$, has the equations of motion (\ref{cond-1}) and (\ref{PlebConn}), and therefore it describes GR with or without a cosmological constant $\Lambda$. 

For the special case of the choice $\alpha_2=0$, Eq. (\ref{fquadratic}) is a linear function in ${\rm Tr}\Psi$ which, together with the action (\ref{Action}), leads to 
\begin{eqnarray}
S_{\rm{linear}}[A^i,\rho]=\alpha_1 \int \rho  (2 {\rm Tr}X-\Lambda), \label{slinear}
\end{eqnarray}
where we use $\Psi=X-(1/3)\Lambda$. This action, which also works well for both $\Lambda=0$ and $\Lambda\neq0$, corresponds to the simplest case of an action belonging to our class of formulations. Notice that (\ref{slinear}) has the same form of the action principle (3) of Ref.~\cite{erratum}. It is important to point out that the action principle (\ref{slinear}) and the one of Ref.~\cite{erratum} were obtained following different approaches. In fact, Eq. (3) in Ref.~\cite{erratum} was obtained from the Plebanski action. In the case $\alpha_2\neq0$, the quadratic term in (\ref{fquadratic}) remains and gives a different action principle with the same equations of motion of the case $\alpha_2=0$. Moreover, the action with the quadratic term has no analog with any other gauge connection formulation for GR. 

Note that $\alpha_2$ is an  arbitrary parameter for $\Lambda=0$ and $\Lambda\neq0$ (except $\alpha_2\neq-\alpha_1/\Lambda$),  and that appears in the action principle but not in the equations of motion. Then, in this sense, $\alpha_2$ resembles the Barbero-Immirzi parameter involved in the Holst action~\cite{holst}. On the other hand, it can be shown that there are not polynomial functions on ${\rm Tr}\Psi$ with degree three or greater satisfying conditions (i) and (ii)  either with $\Lambda=0$ or with $\Lambda\neq0$. Then, the quadratic function (\ref{fquadratic}), under the mentioned conditions on $\alpha_1$ and $\alpha_2$, is the most general case for a viable polynomial function on ${\rm Tr}\Psi$ for GR. 

\section{A method to obtain $\pmb{f({\rm Tr}\Psi)}$}
In addition to the quadratic function, there are many other functions satisfying conditions (i) and (ii). However, instead of giving another example, we shall briefly present  a method to obtain such viable functions for GR, {\it i.e.}, a method to construct action principles belonging to our class of formulations (\ref{Action}). 

Consider a holomorphic function $h({\rm Tr}\Psi)$ in the  domain $\Omega$, which has only a zero of order $\kappa$ at ${\rm Tr}\Psi=0$. Furthermore, it is well known~\cite{libro} that for such $h$, a holomophic function $q({\rm Tr}\Psi)$ exists in $\Omega$, which has no zeros and such that  $h=({\rm Tr}\Psi)^\kappa q$. This suggests that it is convenient to construct $f$ from $h$ via the complex differential equation
\begin{eqnarray}
f- \frac{1}{2} \left( {\rm Tr}\Psi + \Lambda \right) f' = ({\rm Tr}\Psi)^\kappa q. \label{diffeq}
\end{eqnarray}
The reason to consider this particular equation is that its solution, the function $f$, is holomorphic in $\Omega$ and satisfies condition (i). Indeed, the solution is found to be
\begin{eqnarray}
f=   ({\rm Tr}\Psi+\Lambda)^2 (p+\alpha) , \label{sol-diff-eq}
\end{eqnarray}
where the function $p({\rm Tr}\Psi)$ is a primitive of $-2({\rm Tr}\Psi)^\kappa q/({\rm Tr}\Psi  + \Lambda)^3$ and  $\alpha$ is an arbitrary constant. It can be verified that the term $ ({\rm Tr}\Psi+\Lambda)^2p$ is holomorphic at $-\Lambda$, despite the fact that $p$ is not. It can also be  checked by direct calculation that (\ref{sol-diff-eq}) satisfies condition (i). For instance, the quadratic function (\ref{fquadratic}) arises from the simplest case of $h$, namely, $\kappa=1$ and $q=\alpha_1$~(=const). From this point of view, $\alpha_2=\alpha$ is actually an integration constant that  is not necessarily equal to zero.

Now, it is convenient to express condition (ii) in terms of $({\rm Tr}\Psi)^\kappa q$. Then, using (\ref{sol-diff-eq}) to calculate $f'_0$, condition (ii) becomes
\begin{eqnarray}
\lim\limits_{{\rm Tr}\Psi \rightarrow 0}  \left[  \frac{ ({\rm Tr}\Psi)^\kappa q }{ {\rm Tr}\Psi  + \Lambda} - \left( {\rm Tr}\Psi  +  \Lambda \right)  p  \right] \neq  \alpha\Lambda. \label{condii}
\end{eqnarray}
In particular, if $\Lambda=0$, the only possible case satisfying condition (ii) is $\kappa=1$. In such a case, Eq. (\ref{condii}) reduces to 
$q_0 \equiv q |_{{\rm Tr}\Psi=0}  \neq0$, which is satisfied by assumption. For  $\Lambda\neq 0$, the first term on the left-hand side of (\ref{condii}) vanishes, and therefore condition (ii) is equivalent to $p_0 \equiv p |_{{\rm Tr}\Psi=0}  \neq-\alpha$. In this case all possible values of $\kappa$ are allowed.

At this point the reader may have noticed that there are many suitable functions for GR that can be found by following this approach. For example, consider the function $h=\beta {\rm Tr}\Psi \exp{( 2{\rm Tr}\Psi /\Lambda)}$ with $\beta$ a nonzero constant and $\Lambda\neq0$. This function is clearly holomorphic on $\mathbb{C}$ and has a zero  at ${\rm Tr}\Psi=0$, as it is desirable. Thereby, Eq. (\ref{sol-diff-eq}) constructed using $h$ directly satisfies condition (i). Indeed, Eq. (\ref{sol-diff-eq}) gives the function
\begin{eqnarray}
f =  -  \Lambda \beta \exp(2 {\rm Tr}\Psi/\Lambda) + \alpha ({\rm Tr}\Psi+\Lambda)^2, \label{fexp}
\end{eqnarray}
which satisfies condition (i) provided that $\beta\neq0$. It remains to test whether (\ref{fexp}) satisfies condition (ii). To do this we can proceed in two different ways. One way is by verifying condition (ii) directly using (\ref{fexp}). The second one is through the use of (\ref{condii}) with $h$. From any of these ways it follows that condition (ii) is satisfied if $p_0=-\beta/\Lambda\neq-\alpha$. Then, the upshot of our example is a new action for GR with $\Lambda\neq0$, namely, the action (\ref{Action}) with $f$ given by (\ref{fexp}), provided that $\beta\neq0$ and $\beta\neq\alpha\Lambda$. 

To close the analysis of the method described above we can, alternatively, remark that the class of formulations (\ref{Action}) with $f$ defined by the function (\ref{sol-diff-eq}) describes GR provided that the holomorphic function $h({\rm Tr}\Psi)$ has only one zero at ${\rm Tr}\Psi=0$ and satisfies condition (\ref{condii}).

\section{Pure connection formulation}

For the case $\Lambda\neq0$ of the class of formulations (\ref{Action}) it is possible to eliminate the auxiliary field $\rho$, and then get formulations that depend on the gauge connection only.  The aim of this section is to derive such pure connection formulations. We follow a procedure with the same logic as that of Refs.~\cite{CJD2} and~\cite{CJ3}. That is, the elimination of the field $\rho$ is achieved by using its equation of motion. 

Consider that $\Lambda\neq0$ and that $f$ satisfies conditions (i) and (ii). First, we want to express the equation of motion (\ref{cond-1}) in terms of ${\tilde M}$, ${\tilde \mu}$, and $\Lambda$. To do this we make use of the fact that ${\tilde M}=-2 {\tilde \mu} X^2$ can be solved for $X$, and then we use $\Psi=X-(1/3)\Lambda I$ to get ${\rm Tr}\Psi$. Then the Eq. (\ref{cond-1}) reads
\begin{eqnarray}
\pm \frac{1}{(-2{\tilde \mu})^{1/2} } {\rm Tr}{\tilde M}^{1/2}-\Lambda=0, \label{mov-rho}
\end{eqnarray}
where $\tilde{M}^{1/2}$ is a symmetric square root of $\tilde{M}$, that is, $\tilde{M}=\tilde{M}^{1/2} \tilde{M}^{1/2}$ and $\tilde{M}^{1/2}=(\tilde{M}^{1/2})^T$. Recall that such $\tilde{M}^{1/2}$ exists since $\tilde{M}$ is symmetric and nonsingular~\cite{book}. Now, it is straightforward to solve  (\ref{mov-rho}) for ${\tilde \mu}$,
\begin{eqnarray}
{\tilde \mu} = -\frac{1}{2 \Lambda^2} \left( {\rm Tr}{\tilde M}^{1/2} \right)^2. \label{mu}
\end{eqnarray}
With this solution, it is time to get the pure connection actions.  Substituting (\ref{cond-1}) and (\ref{mu}) into (\ref{Action}), we obtain
\begin{eqnarray}
S[A^i]=-\frac{1}{4} \frac{f'_0}{\Lambda} \int \left( {\rm Tr}{\tilde M}^{1/2} \right)^2 d^4x,\label{Actionpure}
\end{eqnarray}
where we have used $f_0\equiv\left. f \right|_{{\rm Tr}\Psi=0}=\frac{1}{2}\Lambda f'_0$ that comes from  the fact that $f$ satisfies condition (i). The outcome is that all the members with $\Lambda\neq0$ belonging to class (\ref{Action})  reduce to the pure connection action principles (\ref{Actionpure}), which share the same dependence on $A^i$, but the factor in front of the action has a dependence on $\Lambda$ through $f'_0/\Lambda$; thus there is a reminiscence of $f$ on this factor. This $\Lambda$ dependence is different from that of Ref.~\cite{Kras1}. Recall that $f'_0$ does not vanish since $f$ satisfies condition (ii), and that in general it involves the cosmological constant. For instance, if we consider the functions (\ref{fquadratic}) and (\ref{fexp}) the factor is $f'_0=2(\alpha_1+\alpha_2\Lambda)$ and $f'_0=2(-\beta+\alpha \Lambda)$, respectively. Furthermore, we also remark that the procedure employed in the derivation of (\ref{Actionpure}) is different from the one followed in Ref.~\cite{Kras1}.

An advantage of the class (\ref{Action}) is that the resulting actions (\ref{Actionpure}) have a very compact form, in contrast to the case of the pure connection formulation for GR with $\Lambda\neq0$ analyzed in Refs. ~\cite{CJ3} and~\cite{Peldan}. Indeed, in Ref.~\cite{Peldan} it is pointed out that the elimination of the auxiliary field from the formulation of the references aforementioned seems to lead to a ``terrible expression'' for the Lagrangian and that for such a reason that calculation was not completely done.

For the treatment of the $\Lambda=0$ case, it is natural to try to apply the same strategy. However, this time the field $\rho$ cannot be solved from (\ref{cond-1}), which is easy to see by setting $\Lambda=0$ in (\ref{mov-rho}). Hence (\ref{Action}), like the CDJ action~\cite{CJD1}, fails to reduce to pure connection action principles in this case. 

\section{Canonical analysis}

In Ref.~\cite{Dadh} it was shown that the particular action (\ref{slinear}) with $\Lambda\neq0$ leads to the usual constraints of  GR in terms of the Ashtekar variables. In this section, we perform the canonical analysis of the class of formulations presented in Sec. \ref{class}. 

By performing the spacetime decomposition we rewrite the class of actions (\ref{Action}) as
\begin{eqnarray}
	S=\int dt d^3x \tilde{\mathcal{L}}=\int dt d^3x {\tilde \mu} f({\rm Tr}\Psi),\label{Action31}
\end{eqnarray}
and Eq. (\ref{FF-cond}) as
\begin{eqnarray}
	F^{(i}{}_{0a}F^{j)}{}_{bc}\tilde{\eta}^{abc} +2 {\tilde \mu} X^i{}_k X^{jk}=0, \label{FF-cond31}
\end{eqnarray}
where $F^{i}{}_{0a}=\dot{A}^i{}_a-D_aA^i{}_0$ and $F^{i}{}_{ab}=\partial_a A^i{}_b-\partial_b A^i{}_a+\varepsilon^{i}{}_{jk}A^j{}_aA^k{}_b$. Here $D_a$ is the covariant derivative corresponding to the spatial connection $A^i{}_a$ and  $a,b,c=1,2,3$ are spatial indices. We define the momenta conjugate to $A^i{}_a$ as
\begin{eqnarray}
\tilde{\pi}^a_i:=\frac{\partial \tilde{\mathcal{L}}}{\partial \dot{A}^i{}_a}=-\frac{f'}{4}(X^{-1})_{ij} \tilde{B}^{aj},\label{momentum}
\end{eqnarray}
where $\tilde{B}^a_i=F_{ibc}\tilde{\eta}^{abc}$ is the corresponding ``magnetic'' field. We will restrict our analysis to nondegenerate magnetic fields, that is, $\det\tilde{B}:=(1/3!)\varepsilon^{ijk}\underaccent{\tilde}{\eta}{}_{abc} \tilde{B}^a_i  \tilde{B}^b_j \tilde{B}^c_k\neq 0$. This restriction implies that  $\det\tilde{\pi}:=(1/3!)\varepsilon^{ijk}\underaccent{\tilde}{\eta}{}_{abc} \tilde{\pi}^a_i  \tilde{\pi}^b_j \tilde{\pi}^c_k=-(f'/4)^3 \det\tilde{B}/\det X\neq 0$ since $f'_0\neq 0$ and $\det X\neq 0$. The phase space variables are $A^i{}_a$ and $\tilde{\pi}^a_i$, whereas  $A^i{}_0$ and $\tilde{\mu}$ are nondynamical variables, since they have vanishing conjugate momenta. 

The Gauss constraint follows from the fact that there are 
no time derivatives of the variables in the spatial projection of the equation of motion (\ref{mot-con}). The constraint then reads
\begin{eqnarray}
\tilde{\mathcal{G}}_i\equiv D_a \tilde{\pi}^a_i \approx 0. \label{gauss}
\end{eqnarray}

The vector constraint can be obtained from the requirement that $X_{ij}$ is symmetric. Indeed, this constraint turns out to be
\begin{eqnarray}
	\tilde{\mathcal{V}}_b\equiv\tilde{\pi}^a_i F^i{}_{ab}=-\frac{f'}{4}(X^{-1})_{ij} F^i{}_{ab} \tilde{\eta}^{acd} F^j{}_{cd} = 0, \label{vector}
\end{eqnarray}
where it can be checked that $F^i{}_{ab} \tilde{\eta}^{ade} F^j{}_{de}=-F^j{}_{ab} \tilde{\eta}^{ade} F^i{}_{de}$.  Notice that the resulting Gauss and vector constraints of the class of formulations are still the same as those for the usual GR in terms of the Ashtekar variables.

The Hamiltonian constraint is derived from the equation of motion (\ref{cond-1}). Explicitly, using ${\rm Tr}\Psi={\rm Tr}X-\Lambda$ and multiplying by $\det\tilde{\pi}$ we can rewrite (\ref{cond-1}) as
\begin{eqnarray}
\left({\rm Tr}X-\Lambda\right)\det\tilde{\pi} =0.  \label{scalarTr}
\end{eqnarray}
Using now the expression for the momenta $\tilde{\pi}^a_i$, we get
\begin{eqnarray}
X^{ij} \det\tilde{\pi}=-\frac{f'}{8}\varepsilon^{jkl}\underaccent{\tilde}{\eta}{}_{abc} \tilde{\pi}^a_k  \tilde{\pi}^b_l \tilde{B}^{ci}. \label{Xmomenta}
\end{eqnarray}
Taking the trace of (\ref{Xmomenta}) and substituting it into (\ref{scalarTr}) gives the Hamiltonian constraint
\begin{eqnarray}
\tilde{\tilde{\mathcal{H}}}&\equiv& \frac{f'}{8} \varepsilon^{ijk}\underaccent{\tilde}{\eta}{}_{abc} \tilde{\pi}^a_i  \tilde{\pi}^b_j \tilde{B}^c_k + \frac{\Lambda}{6}  \varepsilon^{ijk}\underaccent{\tilde}{\eta}{}_{abc} \tilde{\pi}^a_i  \tilde{\pi}^b_j \tilde{\pi}^c_k \approx 0. \label{hamilconstraint}
\end{eqnarray}
It is interesting to note the presence of function $f'$ in this constraint. The Hamiltonian constraint is then a modification of the familiar Ashtekar version of the Hamiltonian constraint. We recall that $f'$ is a dimensionless holomorphic function of ${\rm Tr}\Psi$ and that $f'_0\neq 0$ since $f$ enjoys the property (ii). For instance, the functions (\ref{fquadratic}) and (\ref{fexp}) lead to $f' =2\alpha_1+2\alpha_2({\rm Tr}\Psi+\Lambda) $ and $f' =  - 2\beta \exp(2 {\rm Tr}\Psi/\Lambda) + 2 \alpha ({\rm Tr}\Psi+\Lambda)$, respectively. In particular, if $\alpha_2=0$ in (\ref{fquadratic}), $f' =2\alpha_1=$ const and hence (\ref{hamilconstraint}) reduces to the usual Hamiltonian constraint for GR, as was first found in Ref.~\cite{Dadh}. 

It is worth mentioning that (\ref{hamilconstraint}) is reminiscent of Krasnov's modified Hamiltonian constraint~\cite{Kras5}, but is not the same. In Ref.~\cite{Kras5} the cosmological constant that appears in the usual Hamiltonian constraint is replaced by an arbitrary function that has a dependence different from $f'$. 

Next, it remains only to consider the Poisson brackets among the constraints. Since the Gauss and vector constraints remain unchanged, the Poisson brackets that must be computed are those that involve the Hamiltonian constraint. Taking into account that $\tilde{\tilde{\mathcal{H}}}$ is gauge invariant, the Poisson bracket of $\tilde{\mathcal{G}}_i$ and  $\tilde{\tilde{\mathcal{H}}}$ does not change. Similarly, the Poisson bracket between  $	\tilde{\mathcal{V}}_a$ and $\tilde{\tilde{\mathcal{H}}}$ gives the known result. Therefore, we are bound to calculate only the Poisson bracket of the Hamiltonian constraint with itself.

Let us introduce the smeared Hamiltonian constraint
\begin{eqnarray}
C_{\underaccent{\tilde}{N}}:=\int d^3x \underaccent{\tilde}{N} \tilde{\tilde{\mathcal{H}}},\label{smearedHamiltonian}
\end{eqnarray}
where the test field $\underaccent{\tilde}{N}$ has weight $-1$. Then, the Poisson bracket of interest is
\begin{eqnarray}
\left\{ C_{\underaccent{\tilde}{N}{}_1},C_{\underaccent{\tilde}{N}{}_2} \right\}\!=\!\int\! d^3x \left( \frac{\delta C_{\underaccent{\tilde}{N}{}_1}}{\delta A^i{}_a }  \frac{\delta C_{\underaccent{\tilde}{N}{}_2}}{\delta \tilde{\pi}^a_i }\! -\! \frac{\delta C_{\underaccent{\tilde}{N}{}_2}}{\delta A^i{}_a }  \frac{\delta C_{\underaccent{\tilde}{N}{}_1}}{\delta \tilde{\pi}^a_i }\right). \label{poissonHamiltonian}
\end{eqnarray}
For the purpose of computing this Poisson bracket, it is convenient to obtain the variation of $C_{\underaccent{\tilde}{N}}$ by using (\ref{Xmomenta}) and $\delta f'=f'' \delta{\rm Tr}\Psi$. Then, 
\begin{eqnarray}
\delta C_{\underaccent{\tilde}{N}}=\int d^3x \underaccent{\tilde}{N} \left[ \frac{1}{4} \gamma_1  \delta \left( \varepsilon^{ijk}\underaccent{\tilde}{\eta}{}_{abc} \tilde{\pi}^a_i  \tilde{\pi}^b_j \tilde{B}^c_k \right) \right. \nonumber\\
+ \left.\frac{1}{6} \gamma_2 \delta\left( \varepsilon^{ijk}\underaccent{\tilde}{\eta}{}_{abc} \tilde{\pi}^a_i  \tilde{\pi}^b_j \tilde{\pi}^c_k \right) \right], \label{variationHamiltonian}
\end{eqnarray}
where the functions $\gamma_1=\gamma_1({\rm Tr}\Psi)$ and $\gamma_2=\gamma_2({\rm Tr}\Psi)$ are given by
\begin{eqnarray}
\gamma_1&:=&\frac{1}{4} \frac{\left(f'\right)^2}{\left(f- \frac{1}{2} \left( {\rm Tr}\Psi + \Lambda \right) f'\right)'},\\
\gamma_2&:=&\Lambda+\frac{1}{2}\frac{\left( {\rm Tr}\Psi + \Lambda \right)^2f''}{\left(f- \frac{1}{2} \left( {\rm Tr}\Psi + \Lambda \right) f'\right)'}.
\end{eqnarray}
Now having (\ref{variationHamiltonian}), we get the required variations
\begin{eqnarray}
\frac{\delta C_{\underaccent{\tilde}{N}}}{\delta A^i{}_a }&=&D_b \left(\underaccent{\tilde}{N} \gamma_1  \varepsilon_i{}^{jk} \tilde{\pi}^a_j \tilde{\pi}^b_k  \right), \label{variationA}\\
\frac{\delta C_{\underaccent{\tilde}{N}}}{\delta \tilde{\pi}^a_i } &=&   \frac{1}{2} \underaccent{\tilde}{N}   \varepsilon^{ijk}\underaccent{\tilde}{\eta}{}_{abc} \tilde{\pi}^b_j (\gamma_1 \tilde{B}^{c}_k +\gamma_2 \tilde{\pi}^{c}_k). \label{variationPi} 
\end{eqnarray}
By substituting (\ref{variationA}) and (\ref{variationPi}) into (\ref{poissonHamiltonian}) the Poisson bracket can finally be written as
\begin{eqnarray}
\left\{ C_{\underaccent{\tilde}{N}{}_1},C_{\underaccent{\tilde}{N}{}_2} \right\}=\int d^3x \gamma_1^2 \underaccent{\tilde}{\underaccent{\tilde}{N}}{}_b \tilde{\tilde{Q}}^{ab} F^i{}_{ac} \tilde{\pi}^c_i, \label{poissonHamiltonian2} 
\end{eqnarray}
where
\begin{eqnarray}
\underaccent{\tilde}{\underaccent{\tilde}{N}}{}_b&:=&\partial_b(\underaccent{\tilde}{N}{}_1) \underaccent{\tilde}{N}{}_2 - \partial_b(\underaccent{\tilde}{N}{}_2) \underaccent{\tilde}{N}{}_1,\\
\tilde{\tilde{Q}}^{ab}&:=&\tilde{\pi}^a_i \tilde{\pi}^b_j \delta^{ij}.
\end{eqnarray}

Therefore (\ref{poissonHamiltonian2}) differs from the usual Poisson bracket only by the nontrivial function $\gamma_1$.  For the functions (\ref{fquadratic}) and  (\ref{fexp}), we obtain, respectively,  
\begin{eqnarray}
\gamma_1=\frac{(\alpha_1+\alpha_2({\rm Tr}\Psi + \Lambda))^2}{\alpha_1}, \label{gamma1cuadratic}
\end{eqnarray}
and
\begin{eqnarray}
\gamma_1=\frac{\Lambda  \left( \beta \exp(2 {\rm Tr}\Psi/\Lambda) - \alpha ({\rm Tr}\Psi+\Lambda) \right)^2}{\beta (2{\rm Tr}\Psi+\Lambda) \exp(2 {\rm Tr}\Psi/\Lambda) }.
\end{eqnarray}
Note that if $\alpha_2=0$ in (\ref{gamma1cuadratic}), i.e., in the case of action principle (\ref{slinear}), we get $\gamma_1=\alpha_1=$ const and hence (\ref{poissonHamiltonian2}) is simply the usual Poisson bracket.

Notice that $\tilde{\mathcal{G}}_i$, $\tilde{\mathcal{V}}_a$, and $\tilde{\tilde{\mathcal{H}}}$ form a set of $3+3+1=7$ first class constraints. Also we have  $3\times3=9$ configuration variables $A^i{}_a$. Therefore, we are left with two complex degrees of freedom per space point, as it should be for complex GR.

\section{CONCLUSIONS}

We conclude with some remarks. (a)~In this paper we have introduced a new class of formulations for GR with either a vanishing or a nonvanishing  cosmological constant that depends on a $SO(3,\mathbb{C})$ gauge connection and a complex-valued 4-form, via a holomorphic function of the trace of a symmetric $3\times3$ matrix $\Psi(A^i,\rho)$ that was constructed from these variables. (b)~As a consequence of our class, we have achieved a very simple action principle for general relativity, given by (\ref{Action}) with (\ref{fquadratic}), that involves a quadratic function on the trace of $\Psi$. One advantage of this action principle is that it works well with and without a cosmological constant, a result that contrasts with the action principles of Refs.~\cite{CJD1} and~\cite{Kras1}, which do not either support or necessarily require a nonvanishing cosmological constant. An interesting novelty of this action is that it involves an arbitrary parameter that resembles the Barbero-Immirzi parameter. Furthermore, in the particular case when this quadratic function is reduced to a linear one, the resulting action principle can be related to the one found in Ref.~\cite{erratum}. (c)~We have also developed a method for constructing action principles for general relativity belonging to the class of formulations by integrating holomorphic functions with certain desirable properties. (d)~In particular, and as a straightforward implementation of the method, we have also reported a new action principle for general relativity with a nonzero cosmological constant, given by (\ref{Action}) with (\ref{fexp}), which involves an exponential function of the trace of $\Psi$. The action principles mentioned in (b) and (d)  are new and had not been reported before. (e)~We have obtained the pure connection action principles from the members of the class with a nonzero cosmological constant, by eliminating the field $\rho$, and also compared the resulting actions with that of Ref.~\cite{Kras1}. (f) We have carried out the canonical analysis of the class of formulations. It was found that the only constraint that gets modified with respect to those of the Ashtekar formulation is the Hamiltonian constraint.  The modification consists in promoting a constant factor to the nontrivial function $f'$. Therefore, each member of the class of formulations leads to a different Hamiltonian constraint, whereas the Gauss and vector constraints remain unchanged. The analysis shows that the members of the class have 2 complex degrees of freedom per space point, as it should be for complex formulations of GR.

Furthermore, a natural way to generalize the class of formulations presented in this paper is in the spirit of Refs.~\cite{bengtsson} and~\cite{Kras3}, i.e., by promoting the holomorphic function from a dependence on the trace of $\Psi$ to a dependence on the three independent scalar invariants of the matrix $\Psi$. Work in this direction is in progress. Another relevant aspect of future work is to linearize the class of formulations (\ref{Action}), which can be done in the sense of Ref.~\cite{Kras4}.

\section*{ACKNOWLEDGMENTS}

We deeply thank Mariano Celada for useful discussions and help. We wish to thank Alejandro Perez and Riccardo Capovilla for useful comments. We also thank an anonymous referee for encouraging us to improve our manuscript. This work was supported in part by CONACyT, M\'exico, Grant No. 167477-F.

\end{document}